# Simulating vibration transmission and comfort in automated driving integrating models of seat, body, postural stabilization and motion perception

Prof. Dr. **R. Happee**, Dr. **R. Desai**, Dr. **G. Papaioannou**,
Intelligent Vehicles, Delft University of Technology, the Netherlands

**Abstract**

To enhance motion comfort in (automated) driving we present biomechanical models and demonstrate their ability to capture vibration transmission from seat to trunk and head. A computationally efficient full body model is presented, able to operate in real time while capturing translational and rotational motion of trunk and head with fore-aft, lateral and vertical seat motion. Sensory integration models are presented predicting motion perception and motion sickness accumulation using the head motion as predicted by biomechanical models.

## 1. Comfort in Automated Driving

Automated driving holds great promise to provide safe, comfortable and sustainable transport. Passenger cars and trucks are expected to achieve very high safety levels, allowing drivers to engage in other activities. Such vehicles can be driven manually when desired or when used outside the design domain of the automation. At the same time, driverless cars, taxis and buses will also enhance mobility for those who are unable to drive.

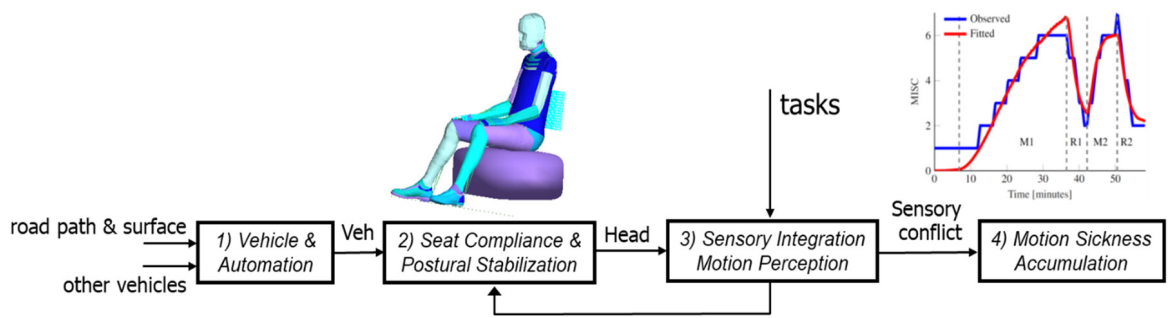

Fig. 1. Comfort simulation using models of: 1) vehicle and automation to predict motion at the seat base, 2) seat and human to predict body and head motion, 3) sensory integration to predict perceived motion used for postural stabilization and sensory conflict leading to 4) motion sickness accumulation in time.

A range of current passenger cars provides *Level 2 automation* [1] through adaptive cruise control and lane-keeping assistance, but these require continuous human supervision to ensure safety. Level 2 automation reduces workload in stop-and-go traffic, but systems occasionally brake for no apparent reason and react abruptly to other road users in cut-in scenarios [2]. Driverless shuttles are tested at low speeds in mixed traffic. In these systems too, abrupt braking in response to other road users is observed. Shuttles are tested on public roads, mostly with a safety steward on board, and a majority of users would not feel safe without human supervision by means of a control room or steward [3]. This provides challenges in creating automated vehicle control strategies, which are not only functionally safe but also perceived to be intuitive, safe and comfortable.

The HTSM roadmap automotive [4] identifies comfort as a main driver for higher levels of automated driving. It defines comfort as an enabler of the *user's freedom for other activities when automated systems are active.* This calls for new vehicle interiors facilitating non-driving tasks and automated driving styles that enhance motion comfort. Hence, **comfort will be pivotal in the acceptance of automated driving**. We need to resolve or minimise discomfort and create a positive perception of using automation. We need to **reinvent the *"Freude am Fahren"*** and create vehicles in which users can effectively work or relax.

Comfort is a highly complex field involving a range of factors such as motion, visual context, workspace ergonomics, posture, seat pressure, sound and climate [5]. The high comfort levels needed in automated driving will require multiple comfort factors to be well addressed. A particular concern is *self-driving car sickness* [6]. While active drivers rarely suffer from car sickness [7], sickness is reported by two thirds of passengers, in particular when taking their eyes off the road [8]. Motion comfort in automated driving is a rapidly emerging research field. First publications confirm the effects of driving style to be significant, but do not provide breakthrough improvements. A deeper understanding is lacking, creating a need to jointly investigate how perceived comfort depends on:
- Vehicle motion - resulting from vehicle path planning, active suspension control and road disturbances [9, 10].
- Visual context - including views outside the window and on computer screens used in non-driving tasks.
- Seating systems - supporting the human body while providing motion freedom to perform various non-driving tasks [11, 12].

Comfort norms like ISO-2631-1 (1997) describe susceptibility to motion sickness and general discomfort as function of motion frequency at the seat. These norms are used with some success to design vehicle control strategies, but current approaches do not take into account the effects of vision and complex vehicle motions, including coupled translation and rotation.

Neither do they address the effects of non-driving tasks, seat interaction and postural stabilization. Advanced theories and models of motion sickness remain to be validated for (automated) driving. The sensory conflict theory [13-15] relates motion sickness to mismatches between perceived and expected motion and between sensory modalities (e.g. between vestibular and visual). The postural instability theory [16] relates motion sickness to disturbed postural stabilization. Both theories are promising to unravel motion sickness causation in automated driving. When using automation, we lack anticipation of upcoming manoeuvres [17]. When taking one's eyes off the road, a lack of world-referenced visual information creates conflicts with vestibular information. Further conflicts emerge due to imperfect integration of translational and rotational vestibular cues. Visual and vestibular motion perception are head-referenced. Hence, we need to take into account and predict the relation between seat and head motion (the seat-to-head transmission or STHT) [18]. Translational motion of the seat results in substantial head rotations (see Figs 3-5). Our research has clarified how such rotations are controlled by vestibular/visual and muscle feedback [19, 20] and our next challenge is to establish their contribution to comfort perception. Seat motion invokes resonances (gain peaks in Figs 3-5) related to seat compliance, postural control and passive spine dynamics. Such resonances are discomforting, but their origin is only partly understood and methods to resolve such resonances are lacking.

To address these challenges, we adopt a **fundamental approach,** developing **models predicting comfort and capturing essential factors in motion perception** (Fig. 1). Building upon our research in human modelling and postural stabilization [17, 19-29] we develop 3D biomechanical models including postural control and sensory integration in order to predict body motion and motion comfort. These models and the underlying understanding will enable the **effective design of vehicle automation for a wide range of driving conditions, non-driving tasks and individual differences**.

## 2. Simulation of postural stabilization of seated vehicle users

Human biomechanical models have been widely used to simulate human motion in vehicles in impact conditions, in vibration (ride comfort), and in dynamic driving. All such applications require models of the seat compliance and the human body including soft tissue compliance, the skeletal system and postural stabilization. In vertical vibration simplified models have been applied successfully but for other conditions 3D multisegment biomechanical models are needed. Such models tend to be complex and computationally demanding when designed for a wide range of applications [19, 27, 30]. Hence we recently developed a computationally efficient full body model with simplified dynamics performing faster than real time when applied with a relatively simple model of the seat (Fig. 2). The efficient models were tuned for vibration transmission using experimental data [31] with 3D translational and rotational motion of trunk and head with fore-aft, lateral and vertical seat motion. Figs 3-5 show that the efficient model provides a similar accuracy as the more computationally demanding active human model.

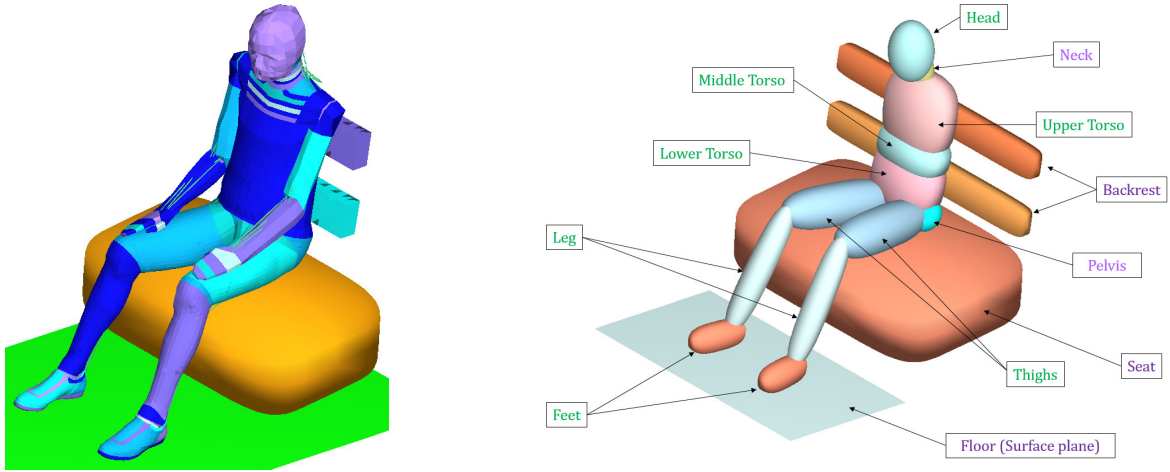

*Fig. 2. Active human model (AHM) for impact and comfort (left) and computationally efficient human model (EHM) for comfort simulation (right) simulating interaction with an experimental seat with configurable backrest [31].*

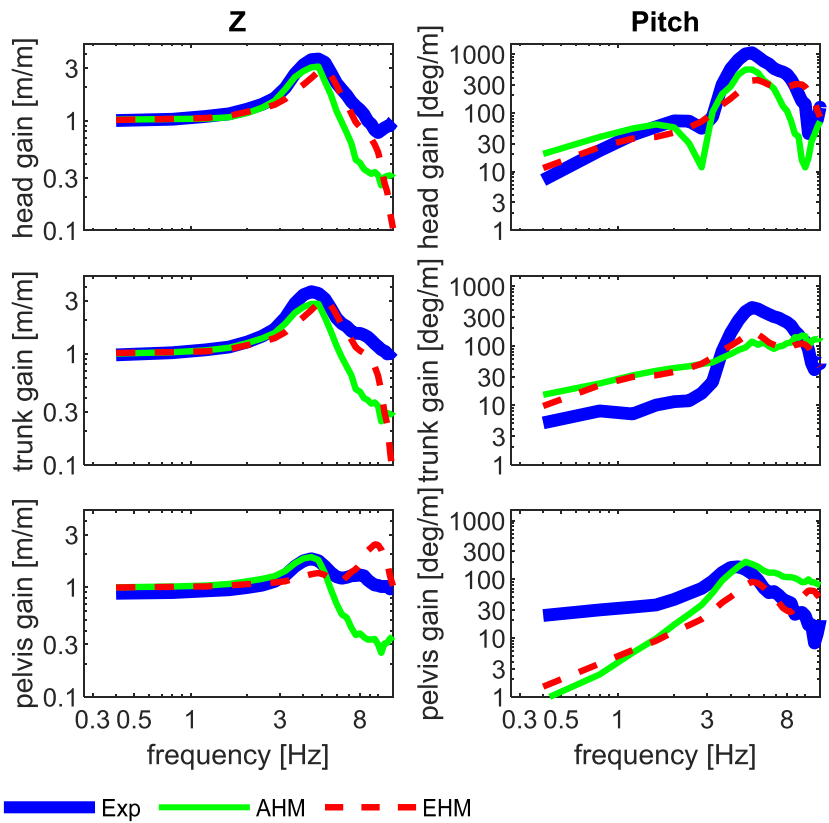

*Fig. 3. Experimental vs. model simulation results in vertical (Z) loading case.*

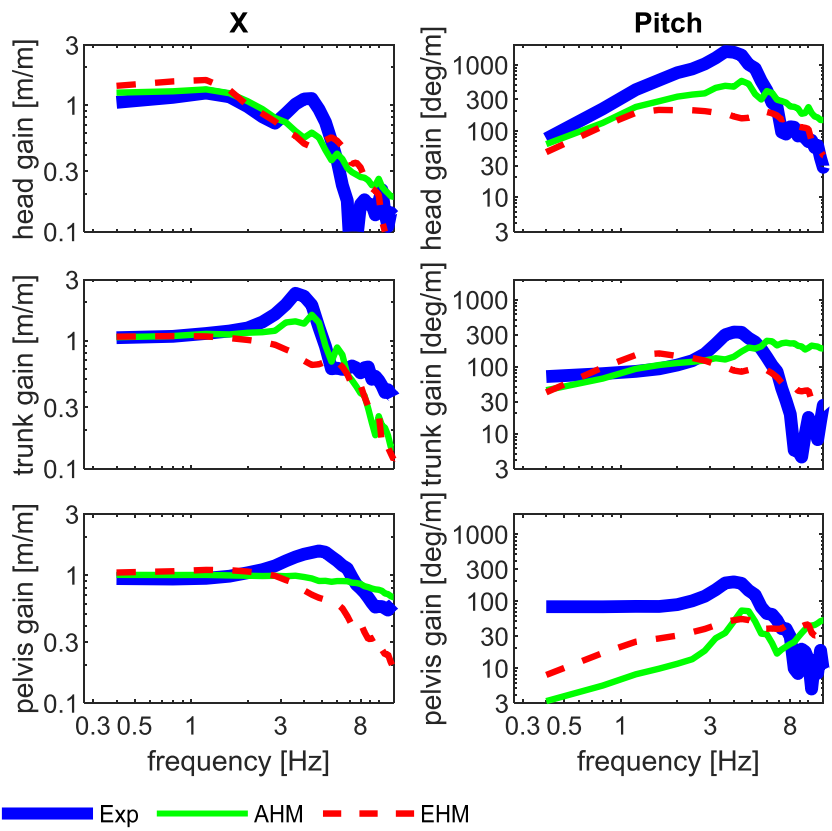

Fig. 4. Experimental vs. model simulation results in fore-aft (X) loading case

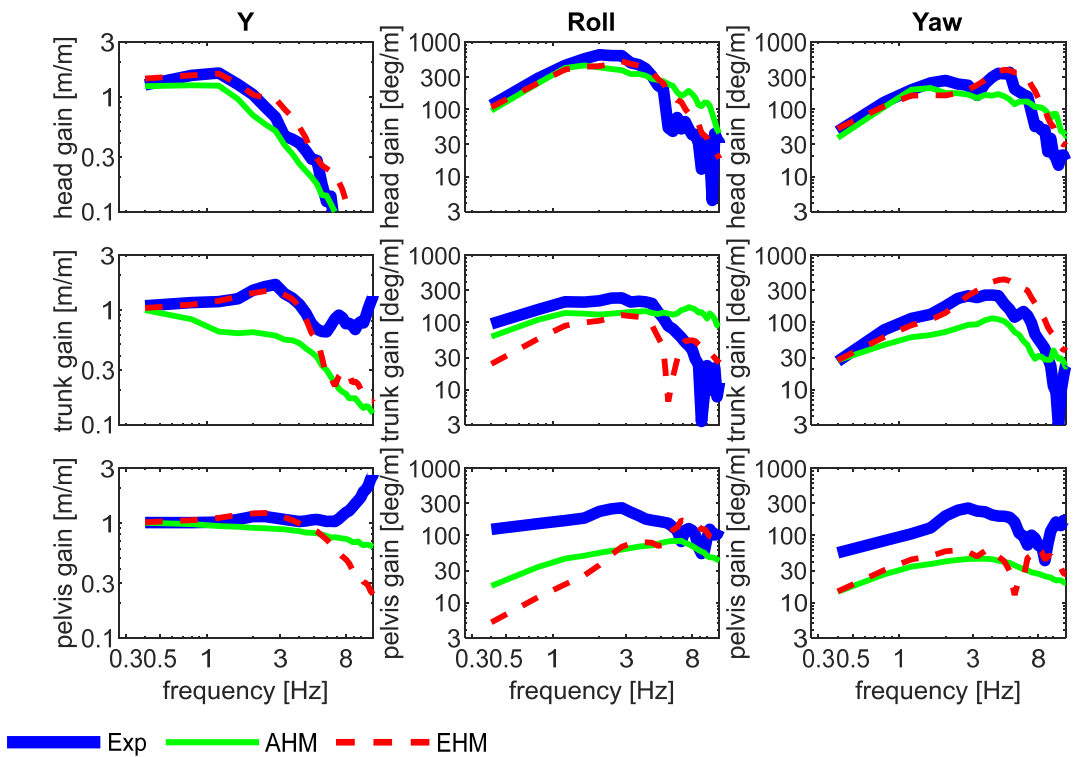

Fig. 5. Experimental vs. model simulation results in lateral (Y) loading case

## 3. Motion Sickness Prediction

Models of sensory conflict have been used to explain and predict motion sickness, motion perception and can be used to capture postural stabilization (Fig. 1). Such models predict conflicts between expected and sensed motion as deriving from vestibular, visual and proprioceptive perception, along with internal models of expected motion in vehicles and other conditions. The predicted conflicts are assumed to lead to accumulation of sickness in time (Fig. 1). We have recently confirmed the predictive capability of such models for motion sickness and motion perception in conditions without vision [32] and with vision [33] in laboratory and test track driving experiments. Recently we adapted such a model to identify individual parameters explaining the substantial differences in sickness susceptibility between participants in varying vision conditions in vehicle experiments and driving simulators [34]. The next challenge will be to extensively validate these sickness prediction models in realistic on road conditions including automated driving [35].